\documentclass[onecolumn,10pt]{revtex4-1}
\usepackage[top=1.0in,bottom=1.0in,left=0.75in,right=0.75in]{geometry}

\usepackage{mathrsfs}
\usepackage{epsfig}
\usepackage{amssymb}
\usepackage{amsfonts}
\usepackage{latexsym}
\usepackage{amsthm}

\usepackage{bm}
\usepackage{graphicx}
\usepackage{lipsum}
\usepackage{amsmath}
\usepackage{hyperref}
\usepackage{fancyhdr}
\usepackage{parskip}
\usepackage{setspace}
\begin{document}

\begin{center}
{\Large \bf The road to precision: Extraction of the specific shear viscosity of the quark-gluon plasma}
\medskip

{\large Chun Shen and Ulrich Heinz \\}
{\it Department of Physics, The Ohio State University, Columbus, OH 43210-1117, USA}
\medskip
\end{center}

\begin{center}
\medskip
{\bf \large Overview}
\medskip
\end{center}

\begin{figure}[b]
  \centering
  \includegraphics[width=0.6\linewidth]{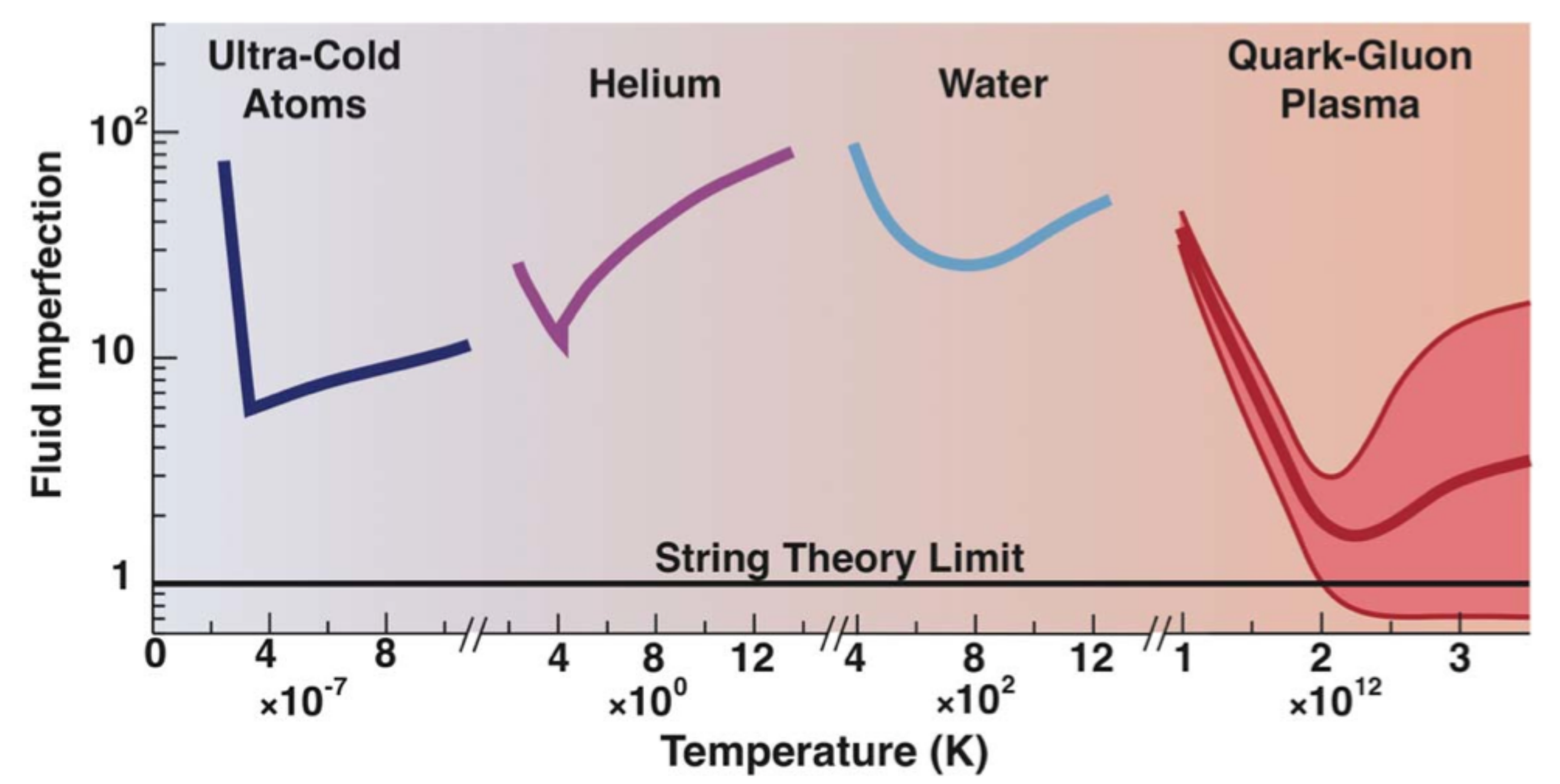}
  \caption{The fluid imperfection index $4\pi \frac{k_B}{\hbar} \frac{\eta}{s}$ of various fluids as a function of temperature. This picture is taken from Tribble R (chair), Burrows A {\it et al.} 2013 {\sl Implementing the 2007 Long Range Plan}, Report to the Nuclear Science Advisory Committee, January 31, 2013. Available at \url{http://science.energy.gov/np/nsac/reports/ .}}
  \label{fig1}
\end{figure}

For a few fleeting moments after the Big Bang the universe was filled with an astonishingly hot and dense soup known as the Quark-Gluon Plasma (QGP), a precursor to the matter we observe today that consisted of elementary particles. Although this Quark-Gluon Plasma also contained leptons and weak gauge bosons, its transport properties were dominated by the strong interaction between quarks and gluons. Utilizing the most powerful particle accelerators, physicists now conduct head-on collisions between heavy ions, such as gold or lead nuclei, to recreate conditions that existed at the birth of the universe. The most striking discovery in relativistic heavy-ion collisions is that the hot and dense matter created during the collisions behaves like an almost perfect (inviscid) liquid, meaning that it can be characterized by a very small shear viscosity $(\eta)$ to entropy density ($s$) ratio (or, ``specific shear viscosity'')\cite{Adams:2005dq,Back:2004je,Arsene:2004fa,Adcox:2004mh}. It turns out that $\eta/s$ for the QGP is smaller than that of any known substance, including that of superfluid liquid helium. In Fig.~\ref{fig1}, we illustrate schematically the specific shear viscosity $\eta/s$ normalized by $\frac{1}{4\pi} \frac{\hbar}{k_B}$, the minimum bound in a large class of theories with infinitely strong coupling \cite{Policastro:2001yc}, for four different types of fluids. The QGP at high temperature exhibits the smallest value of $\eta/s$ of any fluids occurring in nature. 
%
%

Since the discovery around the turn of the millenium of the QGP and its surprisingly perfect fluidity, strong interest has emerged in both theoretical and experimental work to constrain the transport properties of the QGP in relativistic heavy-ion collisions. The specific shear viscosity characterizes one of the most important transport properties of the QGP.  It is at present impossible to compute this transport coefficient with any sort of precision from first principles. 
%
\begin{figure}[t]
  \centering
  \includegraphics[width=0.62\linewidth]{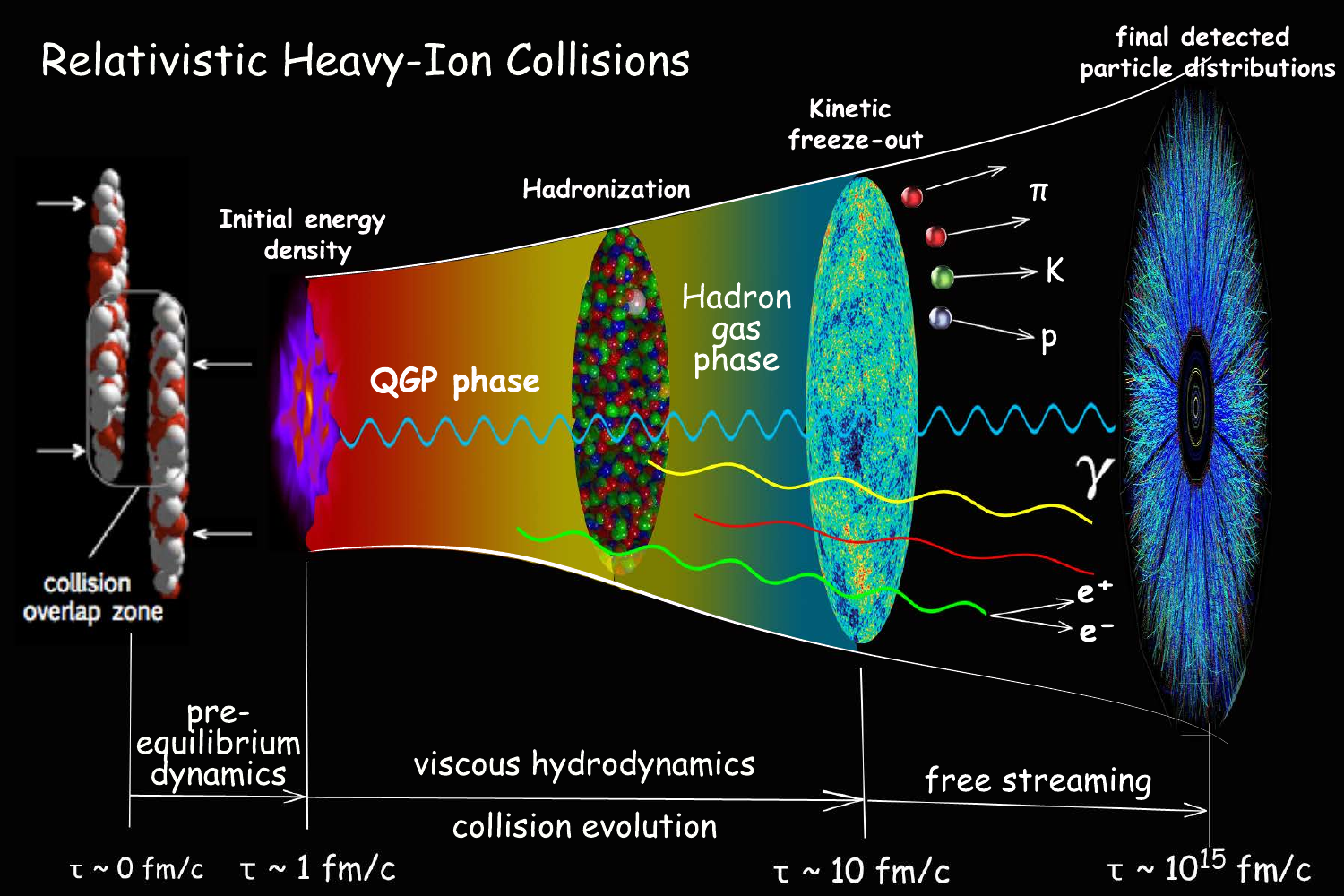}
  \includegraphics[width=0.3\linewidth]{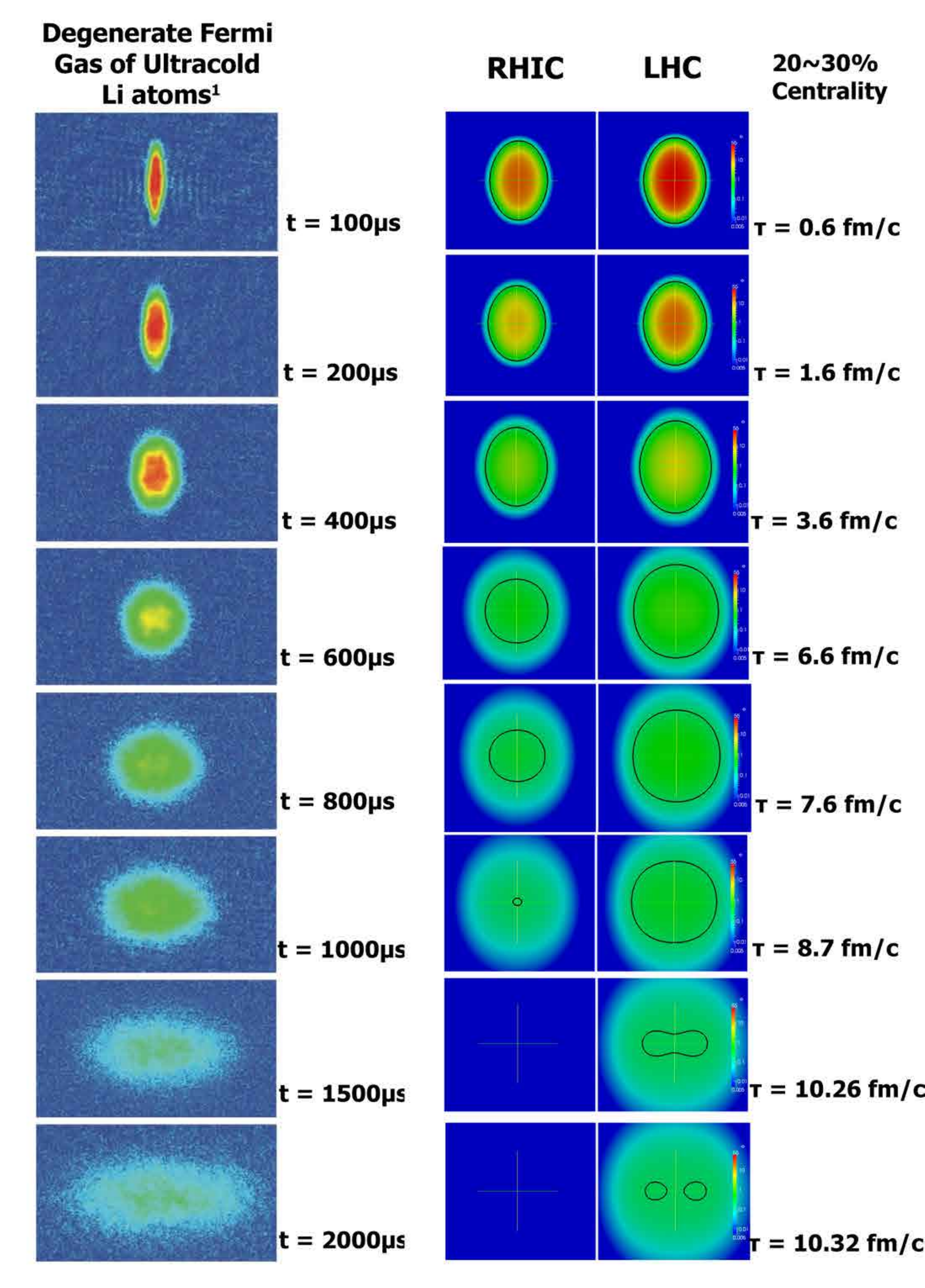}
  \caption{{\it Left Panel:} Illustration of the dynamical evolution of
relativistic heavy-ion collisions. {\it Right Panel:} The collective space time evolution of the collision systems created at the RHIC and the LHC compared with ultra-cold fermi gas \cite{O'Hara:2002zz}. The temperature is color coded. 
}
  \label{fig2}
\end{figure}
%
Alternatively, theoretical analysis of the experiments conducted at the Relativistic Heavy-Ion Collider (RHIC) and the Large Hadron Collider (LHC) offers opportunities to unravel phenomenologically the collective behavior of the QGP created at high energies. Importantly, since the fireballs created in these ``Little Bangs'' are minuscule ($V\sim10^{-42}m^3$) and cool almost instantly ($\sim5\times10^{-23}s$), physicists can only study the QGP through its remnants. This raises unique experimental and theoretical challenges. To trace the  information of the detected stable particles back to the QGP phase at the early stages of the collisions, in order to extract the QGP's properties, 
requires a quantitative understanding of the entire evolution of a heavy ion collision, from the formation and thermalization of the QGP to the dynamics of the hadron resonance gas into which it eventually decays. [A schematic view of the evolution as we currently understand it is illustrated in the left panel of Fig.~\ref{fig2}.] Such a reverse engineering process requires good control of all the model uncertainties, in addition to a consistent description of a large variety of measured observables.

In this featured article we highlight the major recent advances in theory and experiment in constraining the QGP shear viscosity. We begin by explaining the main idea how to measure $\eta/s$ in heavy-ion collisions.  

\begin{center}
\medskip
{\bf \large The QGP viscometer}
\medskip
\end{center}

{\noindent \bf $\bullet$ Bulk dynamics and charged hadron elliptic flow: the state of the art}
\smallskip

The success of viscous hydrodynamics in describing the bulk evolution of relativistic heavy-ion collisions at RHIC \cite{Song:2011qa} and predicting and describing flow data later collected at the LHC \cite{Shen:2011eg, Song:2013qma} leads to an important conclusion: At the macroscopic level, initial anisotropic pressure gradients in the fireball drive the system to collectively develop a momentum anisotropy, as illustrated in the right panel of Fig.~\ref{fig2}. This dynamic evolution shares a great deal of similarity with other strongly coupled many-body systems, such as ultra-cold fermi gases \cite{O'Hara:2002zz}. Inhomogeneities and anisotropies in the initial density distribution can be characterized by the {\em spatial eccentricities} $\{\varepsilon_n, \Phi_n \}$ defined by
\begin{equation}
\varepsilon_1 e^{i \Phi_1} = - \frac{\int r dr d\phi\, r^3 e(r, \phi) e^{i \phi}}{\int r dr d\phi\, r^3 e(r, \phi)} \quad \mbox{ and } \quad
\varepsilon_n e^{i n \Phi_n} = - \frac{\int r dr d\phi\, r^n e(r, \phi) e^{i n \phi}}{\int r dr d\phi\, r^n e(r, \phi)} (n \ge 2),
\end{equation}
where $r$ and $\phi$ are polar coordinates in the transverse plane perpendicular to the beam direction.
(At top RHIC and LHC energies, the dynamics along the beam direction is, to good approximation, boost-invariant, i.e. independent of the longitudinal motion of the reference frame, and therefore less informative for the present discussion than the transverse dynamics.) Similarly, the azimuthal dependence of the emitted particles' momentum distribution can be characterized by the {\em anisotropic flow coefficients} $v_n$ and their associated event plane angles $\Psi_n$, defined by
\begin{equation}
v_n\, e^{i n \Psi_n} = \frac{\int d \phi_p\, \frac{dN}{dy d\phi_p}\, e^{i n \phi_p}}{\int d \phi_p\, \frac{dN}{dy d\phi_p}} \quad \mbox{ and } \quad v_n(p_T)\, e^{i n \Psi_n(p_T)} = \frac{\int d \phi_p\, \frac{dN}{dy p_T dp_T d\phi_p}\, e^{i n \phi_p}}{\int d \phi_p\, \frac{dN}{dy p_T dp_T d\phi_p}},
\end{equation}
where $p_T$ and $\phi_p$ are polar coordinates in transverse momentum space. The conversion efficiency from the initial $\{\varepsilon_n\}$ to the final $\{v_n\}$ is degraded by viscosity in the medium \cite{Song:2010mg}; this is the main effect that opens the door to measuring $\eta/s$ with anisotropic flow observables.

\begin{figure}[t]
  \centering
  \includegraphics[width=0.6\linewidth]{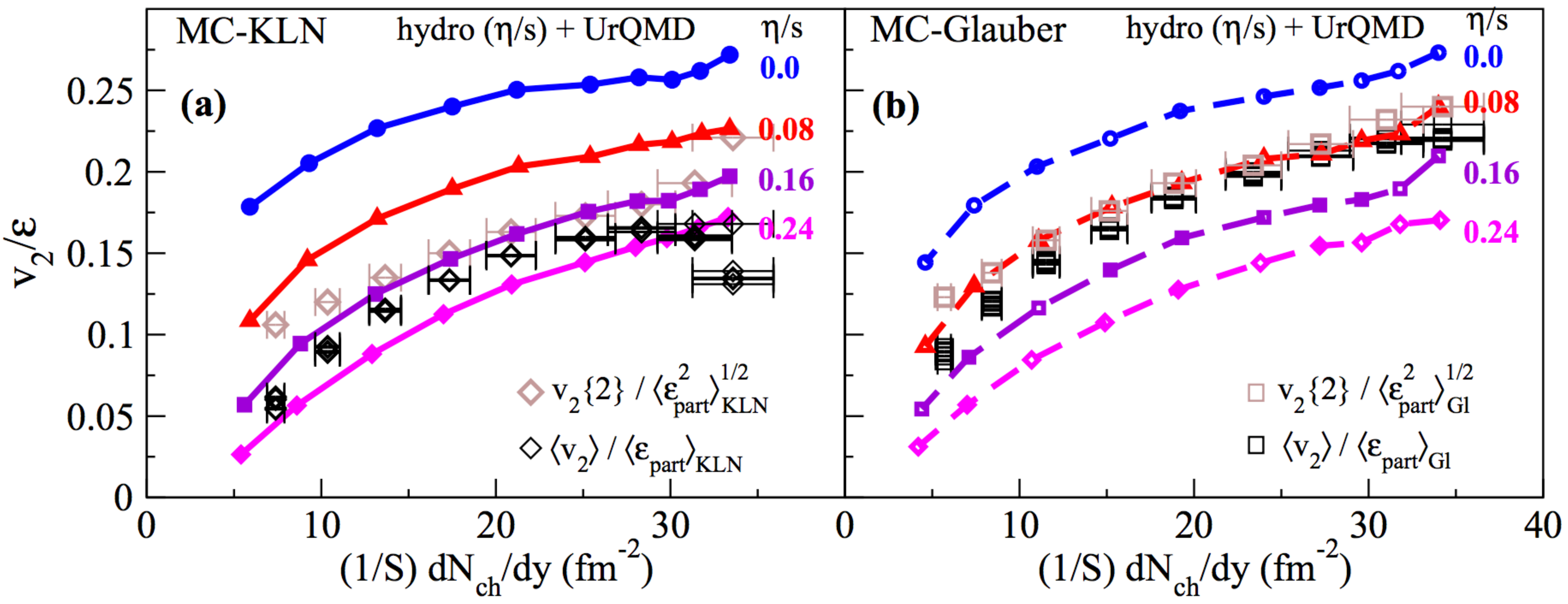}
  \caption{Eccentricity-scaled charged hadron elliptic flow as a function of final charged hadron multiplicity per overlapping area compared to a viscous hydrodynamics + hadron cascade hybrid simulation in Au+Au collisions at top RHIC energy. The plot is taken from \cite{Song:2010mg}. }
  \label{fig3}
\end{figure}

In particular, the second order anisotropic flow coefficient, dubbed the elliptic flow $v_2$, of charged hadrons was first measured in heavy-ion experiments at the CERN SPS and RHIC. In Fig.~\ref{fig3}, we compare the RHIC measurements to results from viscous hydrodynamic simulations, in order to extract $\eta/s$ of the QGP. By confronting state-of-the-art phenomenological modeling with precise experimental measurements, the specific shear viscosity of the QGP can be constrained as \cite{Song:2010mg}
\begin{equation}
\frac{1}{4\pi} \le \left(\frac{\eta}{s}\right)_\mathrm{QGP} \le 2.5 \times\frac{1}{4\pi}.
\end{equation}
A glance at Fig.~\ref{fig1} shows that even the upper limit lies much below the minimal specific shear viscosities seen observed in any other previously known fluid. However, since the central value is so small, the remaining relative uncertainty is still large. It is dominated by model uncertainties in the initial conditions (the two panels in Fig.~\ref{fig3} show the same data compared with two typical initial-state models); it is irreducible unless further anisotropic flow observables, in addition to the elliptic flow, are taken into account.
%

\smallskip
{\noindent \bf $\bullet$  Event-by-event anisotropic flow distributions: disentangling the influence of viscosity and initial state fluctuations}
\smallskip

With improved detection techniques at RHIC and new data at the higher LHC energies (where each collision creates more than twice as many particles as at RHIC), higher order anisotropic flow coefficients of particle momentum distributions have now also been precisely measured \cite{ALICE:2011ab}. The full anisotropic flow spectrum $\{v_n\}$ provides us with enough information to disentangle the extraction of 
%
\begin{figure}[h]
  \centering
  \includegraphics[width=0.28\linewidth]{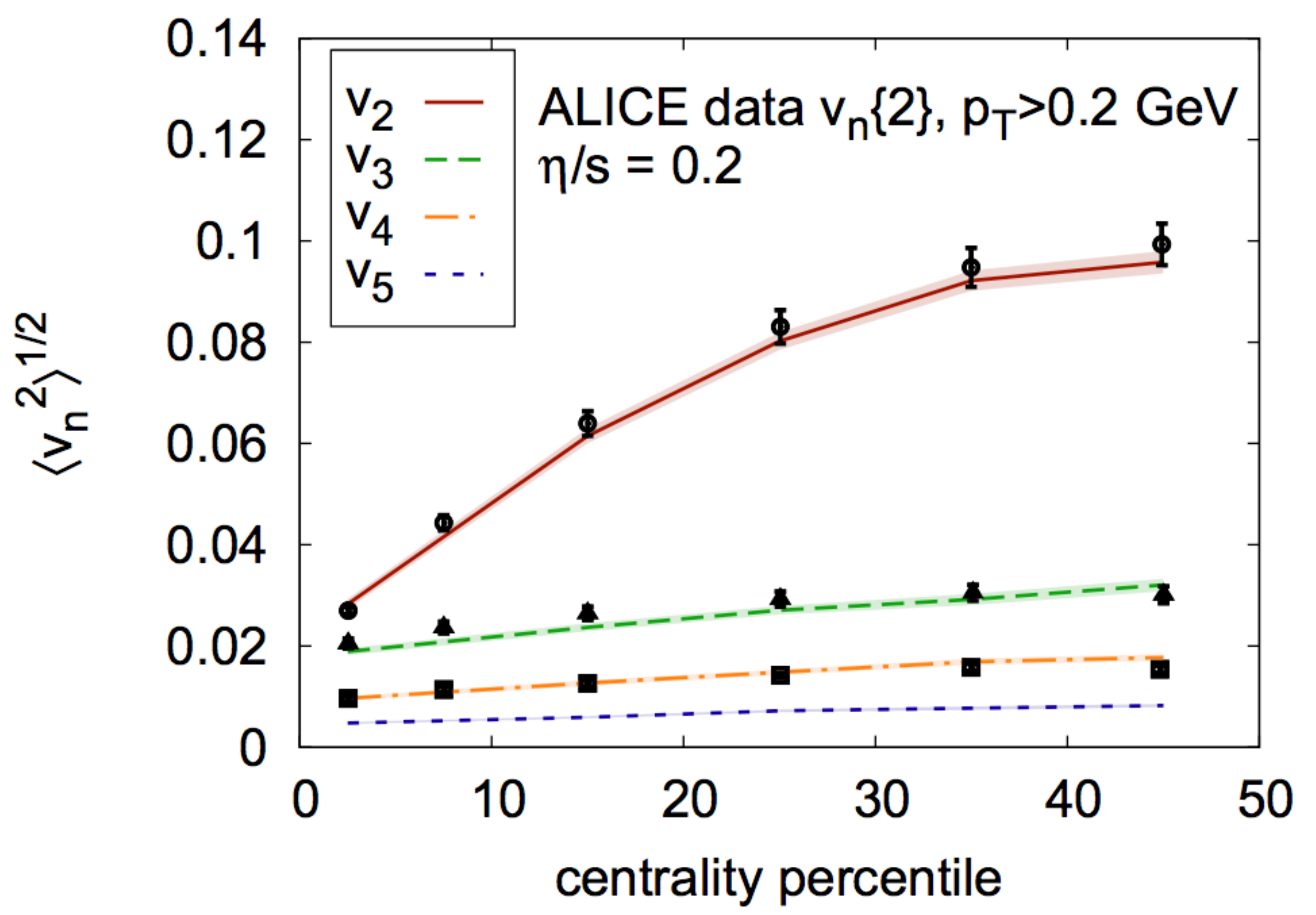}
  \includegraphics[width=0.35\linewidth]{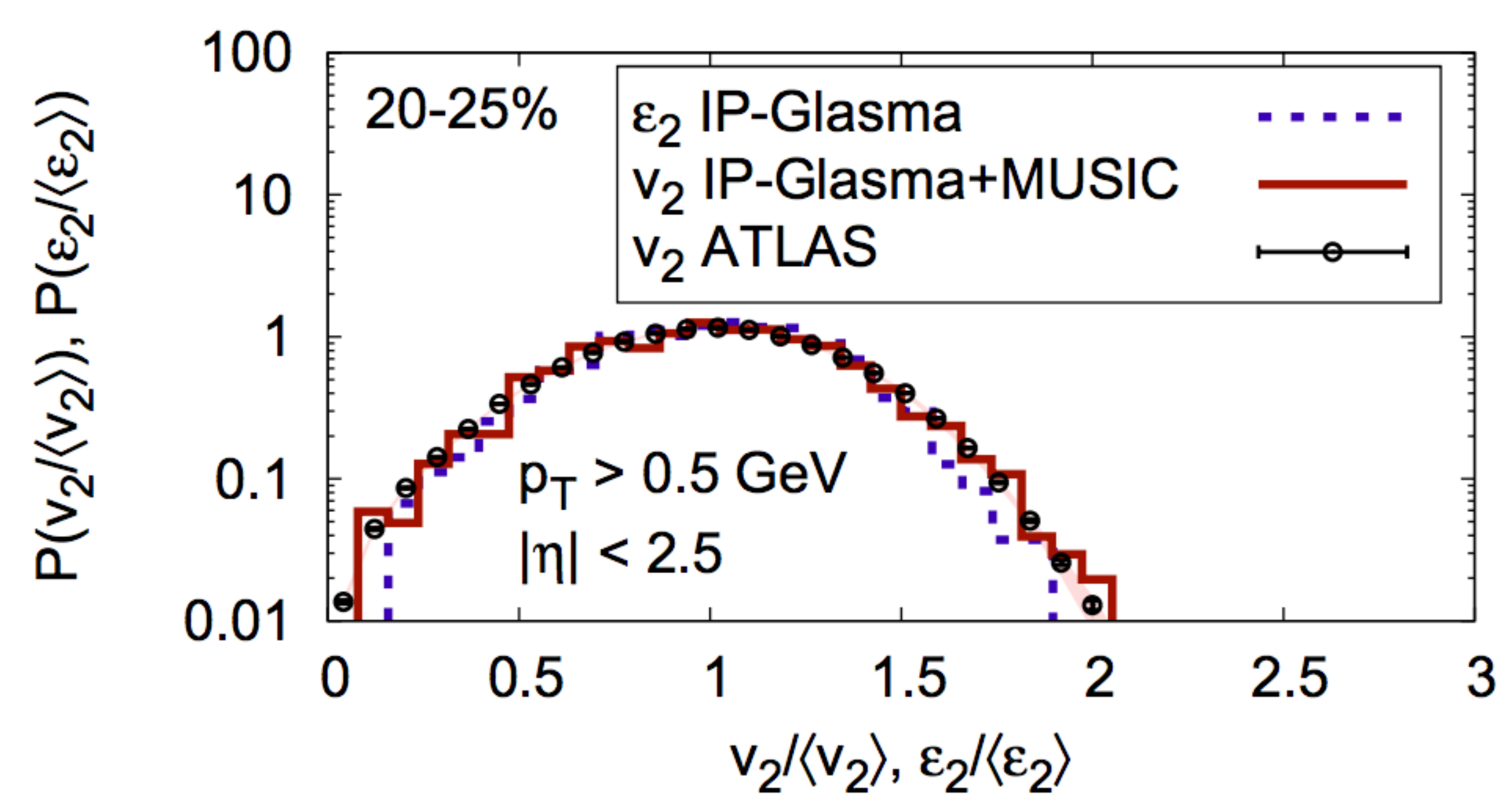}
  \includegraphics[width=0.35\linewidth]{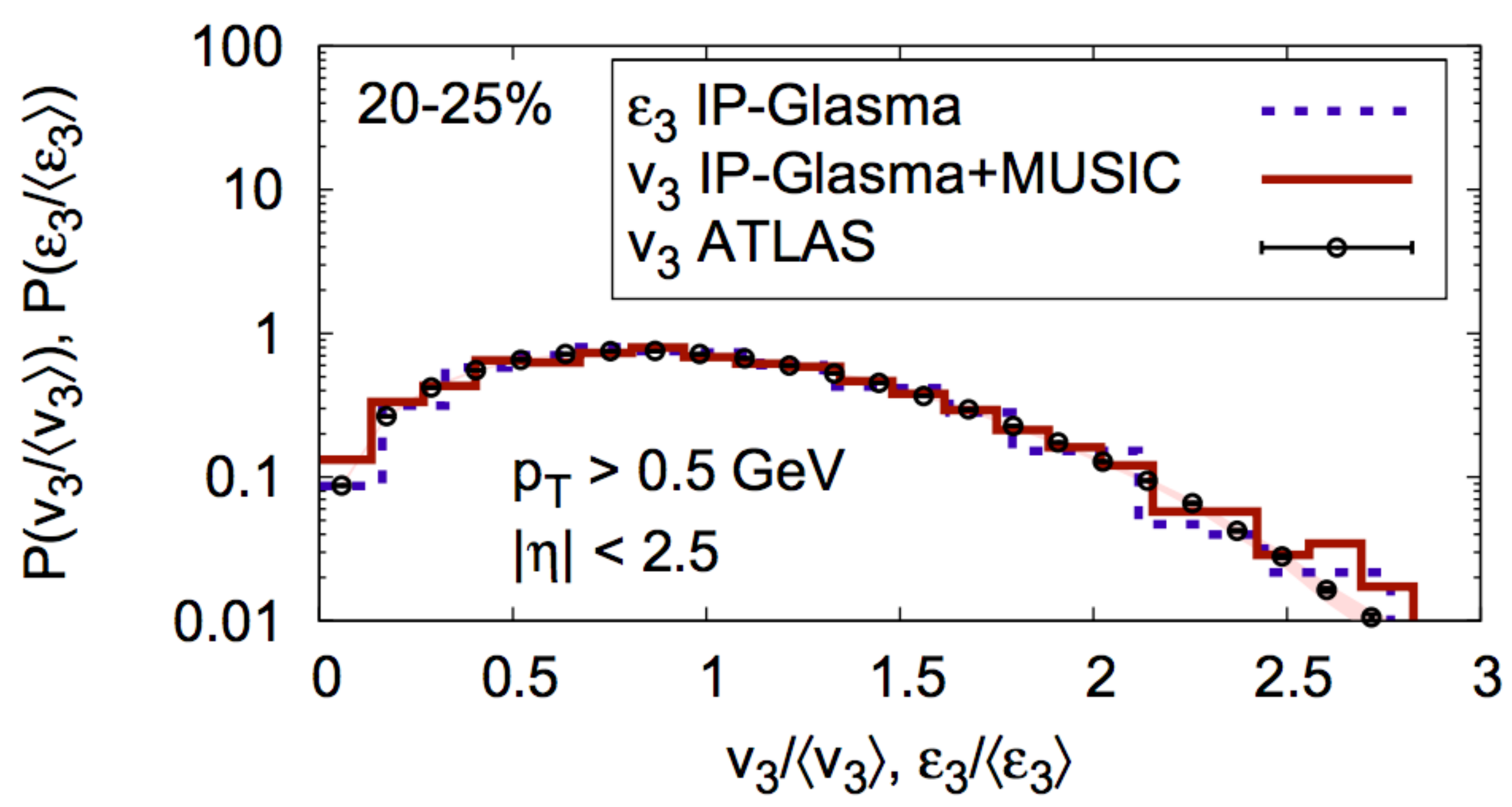}
  \caption{{\it Left Panel:} Charged hadrons anisotropic flow $v_n$ compared with the state-of-the-art IP-Glasma + viscous hydrodynamic (MUSIC) simulations for Pb+Pb at 2.76 $A$ TeV. {\it Middle and right panels:} Event-by-event $v_{2,3}$ probability distributions compared with the same model simulations. The plots are taken from \cite{Gale:2012rq}.}
  \label{fig4}
\end{figure}
%
$\eta/s$ for the QGP from the dynamical consequences of the geometric shape of the initial-state and its quantum fluctuations that were at the heart of the model uncertainty seen in Fig.~\ref{fig3}  \cite{Qiu:2011hf,Luzum:2013yya}. Figure~\ref{fig4} shows recent results from (3+1)-dimensional viscous fluid dynamical simulations \cite{Schenke:2010rr} performed with the currently most advanced initial state model for heavy-ion collisions, the IP-Glasma model \cite{Schenke:2012wb}. The left panel illustrates that this approach provides a precise and consistent description of the measured charged hadron anisotropic flow coefficients, $v_2$ to $v_4$, with $(\eta/s)_\mathrm{QGP} = 0.2$ \cite{Gale:2012rq} (recently refined to $(\eta/s)_\mathrm{QGP}=0.18$ \cite{Schenke:2014zha}). 
The middle and right panels of Fig.~\ref{fig4} show that, even more impressively, the IP-Glasma model initial density distributions, after viscous hydrodynamic evolution, yield an almost perfect description of
the experimentally unfolded $v_n$ probability distributions \cite{Gale:2012rq}. Experimental access to the full spectrum of $v_n$ coefficients and to the probability distributions of their event-by-event fluctuations has changed the scope for a precise determination of the QGP specific shear viscosity: a phenomenological extraction of $(\eta/s)_\mathrm{QGP}$ with a relative precision of order 5-10\% now appears within reach. The same data, as well as several additional recently identified flow fluctuation and flow correlation observables that focus more specifically on the fluctuating behaviour of the flow angles $\Psi_n$ \cite{Gardim:2012im,Luzum:2012da,Heinz:2013bua,Bhalerao:2013ina} also promise a way to simultaneously constrain the spectrum of the initial-state quantum fluctuations. The procedure involves extensive modeling combined with advanced statistical analysis tools
\cite{Novak:2013bqa}, mirroring modern studies of the cosmic microwave background \cite{Ade:2013kta}.

\begin{figure}[h]
  \centering
  \includegraphics[width=0.9\linewidth]{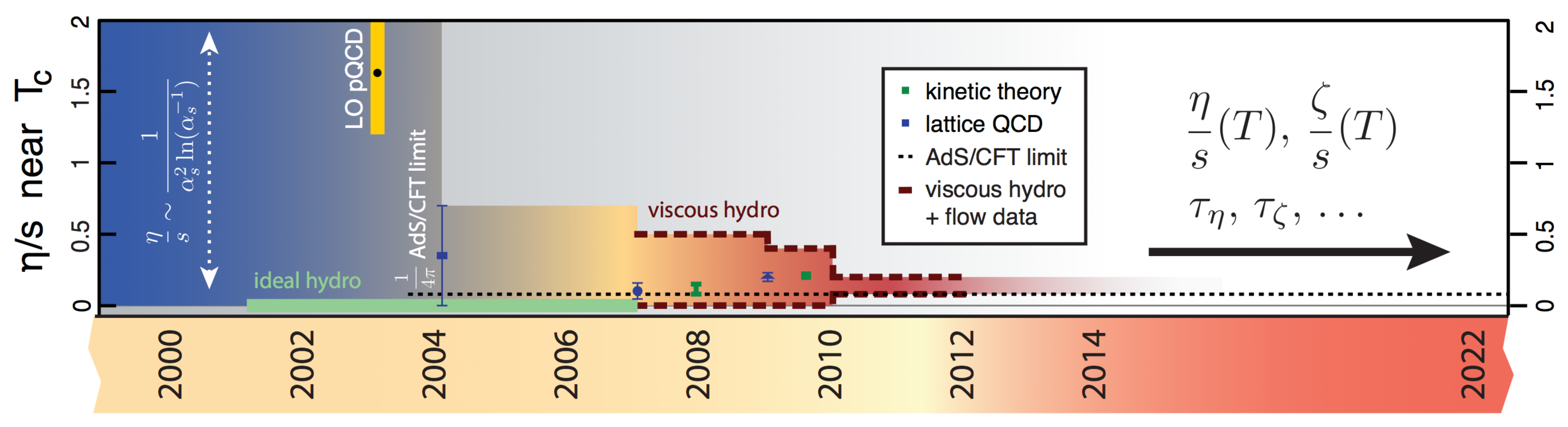}
  \caption{Time line of precise extraction of the specific shear viscosity of QGP. The plot is taken from the ``Hot \& Dense QCD White Paper'', solicited by the NSAC subcommittee on Nuclear Physics funding in the US. Available at \url{http://www.bnl.gov/npp/docs/Bass_RHI_WP_final.pdf}. }
  \label{fig5}
\end{figure}
%
\medskip

In all, through efforts in the development of sophisticated phenomenological models and improvements in the precision of the anisotropic flows measurements, the extraction of the QGP $\eta/s$ has become increasingly accurate. Fig.~\ref{fig5} summarizes a time line of the improvement in constraints placed upon $\eta/s$ of the QGP over the past decade. We can see that the uncertainty of the value of the QGP $\eta/s$ has shrunk dramatically. The convergence is due to progress on both the theoretical and experimental sides.  At the current stage, theoretical models and experimental measurements are both beginning to reach the sensitivity necessary to constrain even the temperature dependence of $(\eta/s)(T)$ \cite{Niemi:2011ix,Shen:2011kn,Shen:2011eg}, as well as that of other transport coefficients, such as the bulk viscosity \cite{Noronha-Hostler:2013gga} and various second-order transport coefficients. 

\begin{center}
\medskip
{\bf \large Outlook and Challenges}
\medskip
\end{center}

\begin{figure}[t]
  \centering
  \includegraphics[width=0.285\linewidth]{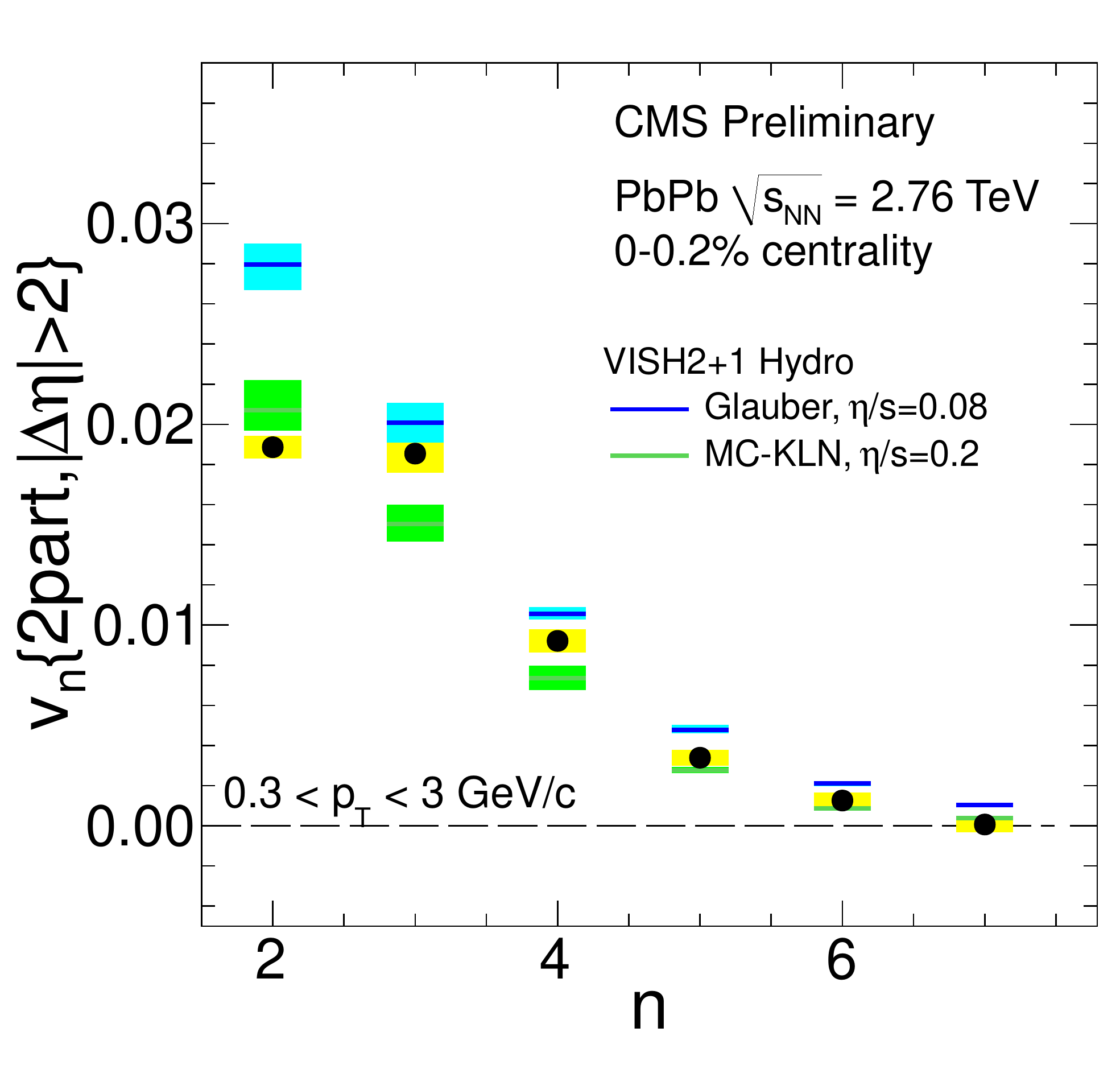}
  \includegraphics[width=0.35\linewidth]{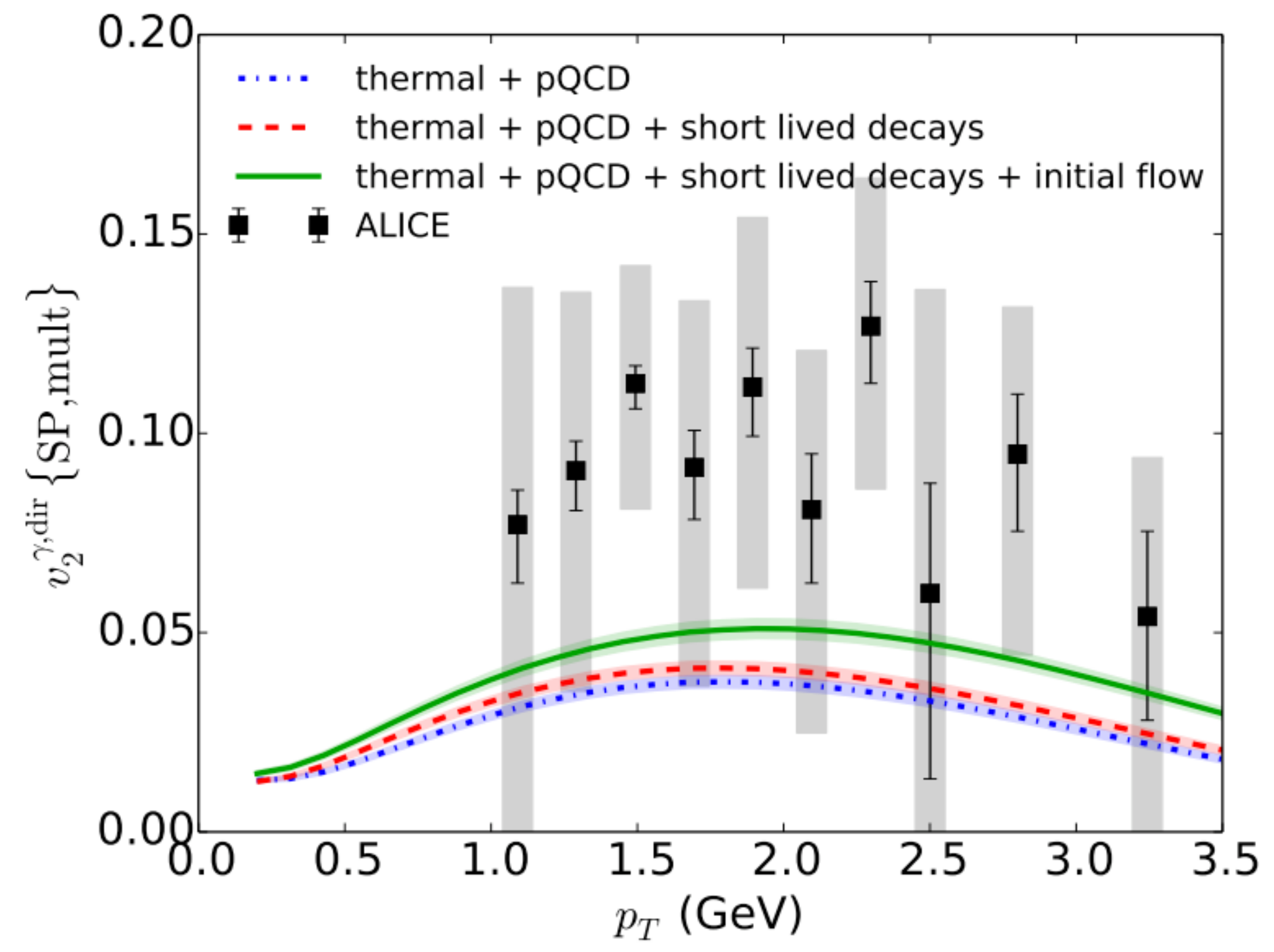}
  \includegraphics[width=0.35\linewidth]{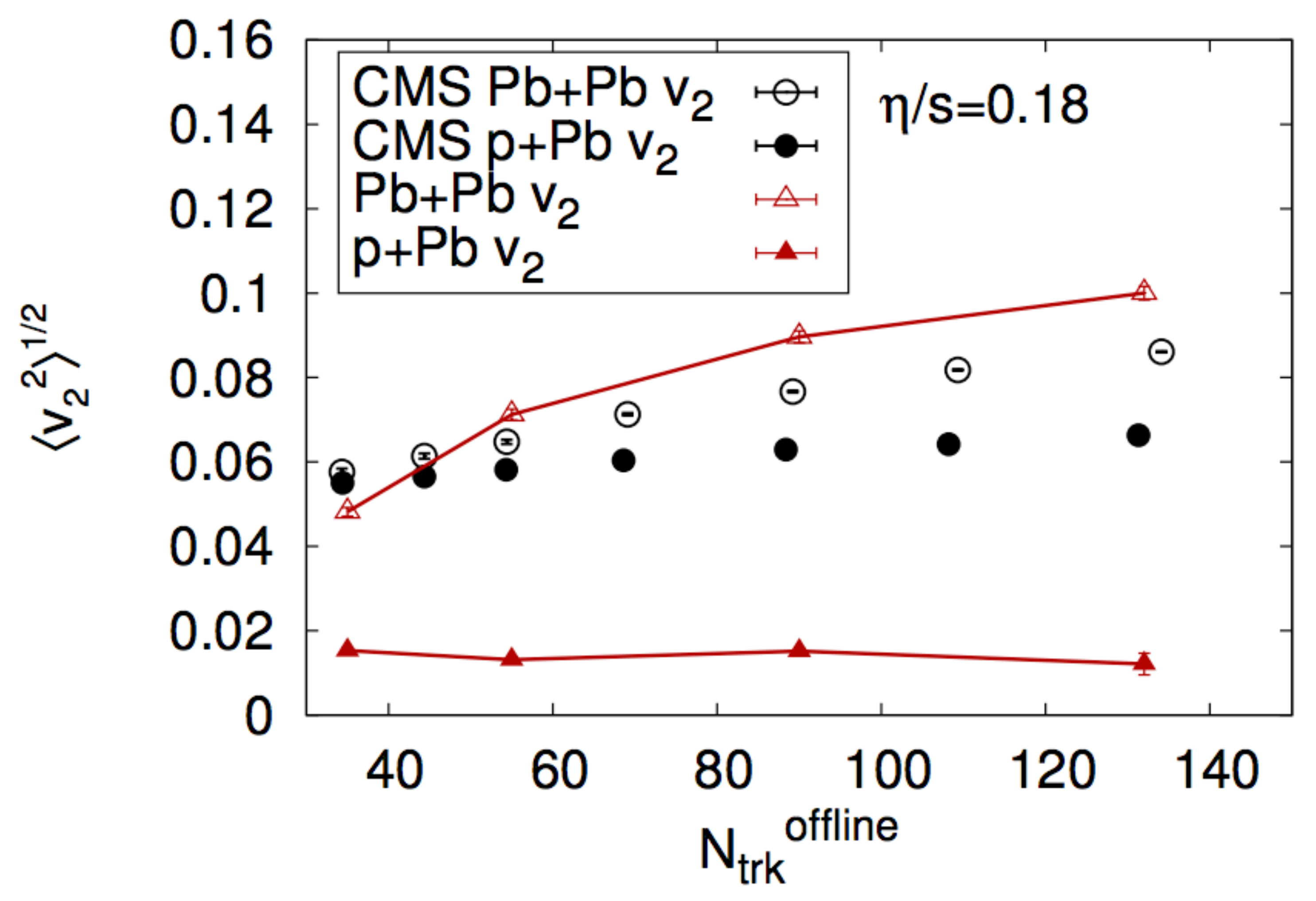}
  \caption{{\it Left Panel:} Charged hadrons anisotropic flow $v_n$ compared with viscous hydrodynamic simulations for 0-0.2\% ultra-central Pb+Pb collisions at 2.76 $A$ TeV \cite{CMS:2013bza}. {\it Middle panel:} Direct photon elliptic flow compared with event-by-event viscous hydrodynamic models \cite{Shen:2014lpa}. {\it Right panel:} Charged hadrons elliptic flow in peripheral Pb+Pb collisions and central p+Pb collisions at LHC compared to the model calculations \cite{Schenke:2014gaa}.}
  \label{fig6}
\end{figure}

While phenomenological modeling with hydrodynamics has led to an excellent description and to 
successful predictions of most soft ($p_T{\,<\,}2$\,GeV/$c$) hadronic observables in relativistic heavy-ion collisions, there are some ``outliers'' that the current state-of-the-art theoretical approaches have difficulties explaining. 

In the left panel of Fig.~\ref{fig6}, the CMS collaboration measured charged hadron anisotropic flow $v_n$ in 0-0.2\% ultra-central Pb+Pb collisions at 2.76 $A$ TeV \cite{CMS:2013bza}. The relatively similar size of the measured $v_2$ and $v_3$ coefficients is a striking feature which challenges current phenomenological model descriptions \cite{Rose:2014fba} (while $\varepsilon_2$ and $\varepsilon_3$ are predicted to be similar in these collisions, shear viscous effects should suppress $v_3$ more strongly than $v_2$). Additionally, in the past 3 years, a surprisingly large {\em direct photon elliptic flow} has been reported by the PHENIX collaboration at RHIC \cite{Adare:2008ab,Adare:2011zr} and later confirmed by the ALICE collaboration at the LHC \cite{Wilde:2012wc,Lohner:2012ct}. All current theoretical models underestimate the measured photon momentum distributions and elliptic flow by factors of 2 to 4 \cite{Chatterjee:2014nta,vanHees:2014ida,Linnyk:2013hta,Shen:2013vja,Shen:2013cca} (a typical model/data comparison is shown in the middle panel of Fig.~\ref{fig6}). This major challenge has become known in the field as the ``direct photon flow puzzle''. Finally, the right panel of Fig.~\ref{fig6} shows that the current state-of-the-art phenomenological model with IP-Glasma initial conditions \cite{Gale:2012rq} fails to provide a consistent description of the measured elliptic (and triangular, not shown) flow coefficients in p+Pb and Pb+Pb collisions at the LHC \cite{Schenke:2014zha}. While this does not necessarily invalidate the hydrodynamic paradigm, it at least shows that our theoretical understanding of the internal structure of the proton and of the fireballs it creates when colliding with a large nucleus is still incomplete.

Resolving these puzzles requires further improvements of the theoretical model from the description of initial state fluctuations to more realistic bulk dynamic evolution and to the detailed reconstruction of the particle correlations in the final states. 

With higher statistics accumulated in the experiments, the anisotropic flow of rare probes, like high-$p_T$ charged hadrons ($p_T > 10$ GeV), direct photons, $D$ mesons, $J/\psi$ and etc., are now being or will soon be measured to satisfactory precision. Compared to the bulk soft hadrons, the anisotropies of these rare particles can probe earlier dynamics of relativistic heavy-ion collisions. In principle, they are more sensitive to the temperature dependence of $\eta/s$ in the high temperature region. A combined analysis of these rare observables with the anisotropic flow of charged hadrons at a variety of collisions energies will help us to further constrain our theoretical models and improve the precision of our extraction of the QGP transport coefficients.

\acknowledgments{This work was supported by the U.S. Department of Energy, Office of Science, Office of Nuclear Physics, under Awards No. DE-SC0004286, and (within the framework of the JET Collaboration) DE-SC0004104. }


\end{document}